\documentclass[final,3p,times,twocolumn]{elsarticle}

\usepackage{hyperref}
\usepackage{amsmath}
\usepackage{xspace}
\usepackage{ulem}
\usepackage[pdftex]{color}

\newcommand{\Vbb}{V_{\mathrm{bb}}}
\newcommand{\Vbpw}{V_{\mathrm{BPW}}}
\newcommand{\Gout}{G_{\mathrm{OUT}}}
\newcommand{\Gcsa}{G_{\mathrm{CSA}}}
\newcommand{\Gsf}{G_{\mathrm{SF}}}
\newcommand{\Gro}{G_{\mathrm{RO}}}
\newcommand{\Acs}{A_{\mathrm{CS}}}
\newcommand{\Cd}{C_{\mathrm{D}}}
\newcommand{\Ci}{C_{\mathrm{i}}}
\newcommand{\Cfb}{C_{\mathrm{FB}}}

\journal{Journal of \LaTeX\ Templates}

\begin{document}

\begin{frontmatter}

\title{Design study and spectroscopic performance of SOI pixel detector with a pinned depleted diode structure for X-ray astronomy}

\author[Miyazakiaddress]{Masataka Yukumoto\corref{mycorrespondingauthor}}
\cortext[mycorrespondingauthor]{Corresponding author}
\ead{yukumoto@astro.miyazaki-u.ac.jp}

\author[Miyazakiaddress]{Koji Mori}
\author[Miyazakiaddress]{Ayaki Takeda}
\author[Miyazakiaddress]{Yusuke Nishioka}
\author[Miyazakiaddress]{Syuto Yonemura}
\author[Miyazakiaddress]{Daisuke Izumi}
\author[Miyazakiaddress]{Uzuki Iwakiri}

\author[Kyotoaddress]{Takeshi G. Tsuru}

\author[DSaddress]{Ikuo Kurachi}

\author[Tokyoaddress]{Kouichi Hagino}

\author[KEKaddress]{Yasuo Arai}

\author[TUSaddress]{Takayoshi Kohmura}

\author[Konanaddress]{Takaaki Tanaka}

\author[Miyazakiaddress]{Miraku Kimura}
\author[Miyazakiaddress]{Yuta Fuchita}
\author[Miyazakiaddress]{Taiga Yoshida}

\author[Kyotoaddress]{Tomonori Ikeda}

\address[Miyazakiaddress]{ Department of Applied Physics, Faculty of Engineering, University of Miyazaki, 1-1 Gakuen-Kibanadai-Nishi, Miyazaki, 889-2192, Japan}
\address[Kyotoaddress]{ Department of Physics, Faculty of Science, Kyoto University, Kitashirakawa Oiwake-cho, Sakyo-ku, Kyoto 606-8502, Japan}
\address[DSaddress]{ D$\&$S Inc., 774-3-213 Higashiasakawacho, Hachioji, Tokyo 193-0834, Japan}
\address[Tokyoaddress]{ Department of Physics, University of Tokyo, 7-3-1 Hongo, Bunkyo, Tokyo 113-0033, Japan}
\address[KEKaddress]{ Accelerator Laboratory, High Energy Accelerator Research Organization (KEK), 1-1 Oho, Tsukuba 305-0801, Japan}
\address[TUSaddress]{ Department of Physics and Astronomy, Faculty of Science and Technology, Tokyo University of Science, 2641 Yamazaki, Noda, Chiba 278-8510, Japan}
\address[Konanaddress]{ Department of Physics, Konan University, 8-9-1 Okamoto, Higashinada, Kobe, Hyogo 658-8501, Japan}

\begin{abstract}
We have been developing silicon-on-insulator (SOI) pixel detectors with a pinned
depleted diode (PDD) structure, named ``XRPIX'', for X-ray astronomy. The PDD
structure is formed in a thick p-type substrate, to which high negative voltage is
applied to make it fully depleted. A pinned p-well is introduced at the backside of
the insulator layer to reduce a dark current generation at the Si-SiO$_{2}$
interface and to fix the back-gate voltage of the SOI transistors. An n-well is
further introduced between the p-well and the substrate to make a potential barrier
between them and suppress a leakage current. An optimization study on the n-well
dopant concentration is necessary because a higher dopant concentration could result
in a higher potential barrier but also in a larger sense-node capacitance leading to
a lower spectroscopic performance, and vice versa. Based on a device simulation, we
fabricated five candidate chips having different n-well dopant concentrations. We
successfully found out the best n-well design, which suppressed a large leakage
current and showed satisfactory X-ray spectroscopic performance. Too low and too high
n-well dopant concentration chips showed a large leakage current and degraded X-ray
spectroscopic performance, respectively. We also found that the dependency of X-ray
spectroscopic performance on the n-well dopant concentration can be largely
explained by the difference in sense-node capacitance.

\end{abstract}

\begin{keyword}
X-ray detectors \sep X-ray SOIPIX \sep Monolithic active pixel sensors \sep Silicon on insulator technology
\end{keyword}

\end{frontmatter}


\section{Introduction} \label{sec:intro}
CCD detectors have been the main instruments of past and current operating X-ray observatories for the last three decades \cite{ASCA, Chandra, XMM, Suzaku, MAXI, Hitomi} and will be still actively used for new X-ray observatories \cite{XRISM, MIT}. CCD detectors are a nearly technically complete instrument and show spectroscopic performance close to the Fano limit of silicon detectors. CCD detectors are expected to continue to advance in the direction of large format and fast readout maintaining the high spectroscopic performance. On the other hand, mostly driven by commercial imaging application, CMOS detectors have achieved major advances with the modern silicon processing technology. Compared to CCD detectors, CMOS detectors are characterized by their lower power consumption and ability to be realized as a system large-scale integration by incorporating advanced signal processing circuitry. These characteristics are not only highly advantageous for simplifying instrument systems but also provide with new potentials beyond CCD detectors.

We have been developing a new silicon pixel CMOS detector, called XRPIX, for X-ray
astronomy \cite{tsuru}. XRPIX utilizes silicon-on-insulator (SOI) technology, which
combines a thick high-resistivity silicon sensor layer and a high-speed
low-resistivity silicon circuit layer via a buried oxide (BOX) layer in a single
monolithic detector \cite{arai}. The most distinctive feature of XRPIX compared to
other silicon pixel detectors is that a self-trigger function is incorporated by
taking advantage of CMOS technology \cite{takeda}. This function enables XRPIX to
achieve a time resolution of 10~{\textmu}sec or higher. With this high time
resolution and surrounding veto counters, the anti-coincidence technique can be
utilized to significantly reduce non-X-ray background events due to cosmic rays,
which hampers the detection of faint X-ray emission in the case of CCD detectors \cite{antico}.

The history of the XRPIX development is almost a history of improvement of its
spectroscopic performance \cite{tsuru}. We experienced several important progresses
so far and the latest one was provided by incorporating a Pinned Depleted Diode
(PDD) structure in the sensor layer \cite{kamehama, harada}. Radiation hardness is
also significantly improved as well \cite{hayashida, Hagino}. The PDD structure of
XRPIX is formed in a high-negative-bias applied p-type substrate. There exists a
pinned p-well under the BOX layer and a depleted n-well between the p-well and the
substrate. The n-well acts as a potential barrier to prevent a leakage current from
the p-well to the substrate. The details of the PDD structure are given in the next
section. An optimization study of the n-well dopant concentration is necessary
because higher dopant concentration could result in a higher potential barrier but
also in a larger sense-node capacitance leading to a lower spectroscopic performance,
and vice versa. In addition, the prototype XRPIX chips with the PDD structure used a
rarely available high-resistivity substrate ($>$25~k$\Omega$\,cm) \cite{kamehama,
harada}. The resistivity of substrates affects the back bias voltage ($\Vbb$)
required for full depletion, and $\Vbb$ affects the height of the potential barrier
the n-well makes. Therefore, the resistivity of substrates is also related to the
optimization. In this paper, we report the experimental results of the optimization
study fabricating five chips with different n-well dopant concentrations using
moderate-resistivity substrates ($\leq$10~k$\Omega$\,cm) that are consistently
available in future. Section~\ref{sec:design} describes the details of the PDD
structure. Section~\ref{sec:tcad_fab} gives the simulation study of n-well
dopant concentration and the specification of the five candidate
chips. Section~\ref{sec:exp} presents the basic characteristics and X-ray
spectroscopic performance of the chips. Then, Section~\ref{sec:discussion} discusses
the relation between the sense-node capacitance and the X-ray spectroscopic
performance. Finally, Section~\ref{sec:conculusion} concludes this study.

\section{Device structure and description}\label{sec:design}
\begin{figure}[htbp]
      \begin{center}
        \includegraphics[scale=0.22]{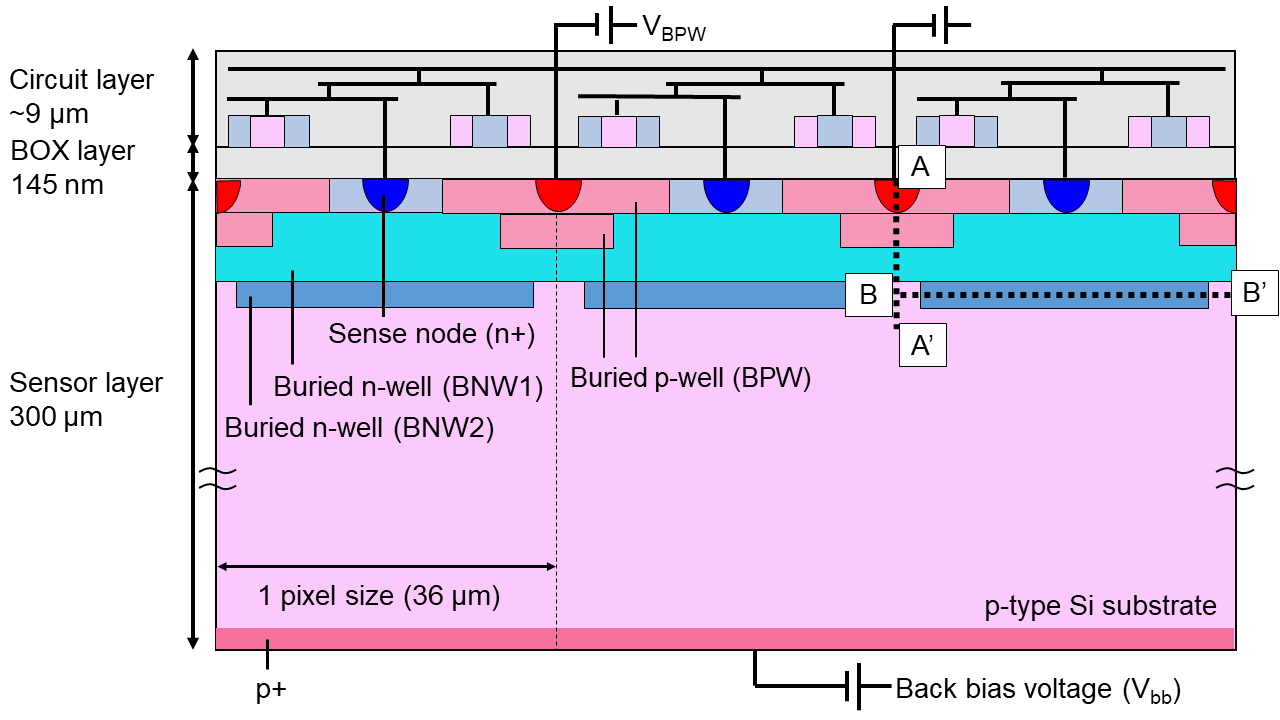}
        \caption{Cross-sectional view of XRPIX8. Two dotted lines, A--A' and B--B', indicate lines, along which potential profiles are drawn in Figures \ref{fig:sim_psijp}(a) and \ref{fig:sim_Psi2}, respectively.}
        \label{fig:schem_str}
      \end{center}
\end{figure}

Figure \ref{fig:schem_str} shows a cross-sectional view of XRPIX8, one of the XRPIX
series with the PDD structure. It consists of three layers: a 9~{\textmu}m thick
circuit layer including five level interconnects and about 40~nm SOI transistors
consisting of low-resistivity silicon, a 300~{\textmu}m thick sensor layer
consisting of high-resistivity silicon, and a 145~nm thick BOX layer between the two
layers. A high negative voltage is applied at the backside of the sensor layer to
fully deplete the substrate. A buried p-well (BPW) is introduced at the backside of
the BOX layer. The neutralized BPW is effective for reducing the dark current
generation at the Si-SiO$_2$ interface between the BOX layer and the sensor layer.
It shares the same concept as that of a pinned photodiode commonly used in CCD and
CMOS image sensors. The BPW, pinned to a fixed bias of $\Vbpw$, also acts as a
shielding layer between the sense node and the circuit layer, preventing extra noise
from transistors. A buried n-well (BNW1) introduced just under the BPW is depleted.
The most important role of the BNW1 is to suppress a large current from the BPW to the p-type substrate. The PN junction with the BPW and BNW1 makes a built-in-potential and it acts as a potential barrier that suppresses the punch-through between the BPW and the p-type substrate.
The BNW1 also forms a lateral electric field to gather carriers generated in the sensor layer into the sense node \cite{kamehama}.
The other buried n-well (BNW2) exists slightly under the BNW1. In the lateral direction, the BNW2 covers only the central portion of a pixel to further enhance the lateral electric field. It is introduced in XRPIX8 for the first time based on the result that the charge collection efficiency degraded near the pixel boundary in the previous prototype XRPIX with the PDD structure \cite{kayama}.

XRPIX8 has 96~$\times$~96 pixels, and the pixel size is
36~{\textmu}m~$\times$~36~{\textmu}m. Pixels at the outer edge of the pixel array
are dummy pixels to suppress the interference from peripheral region. Therefore,
pixels for X-ray detection are 94~$\times$~94 pixels, of which 14~$\times$~94 pixels
are test element groups (TEG) having different circuit configurations.

\section{Simulation study and fabrication of candidate chips} \label{sec:tcad_fab}

\subsection{Simulation study of n-well dopant concentration}\label{sec:tcad}
The potential barrier height depends on the n-well dopant concentration and $\Vbb$.
We made a simulation study on the potential barrier behavior
with different dopant concentrations of BNW1 using the TCAD tools (HyENEXSS\footnote{HyENEXSS is managed by TCAD Academic Council. E-mail address: tac@appi.keio.ac.jp})
\cite{TAC, selete, wada, nakamura}.
Referring to the operating condition of the previous prototype
XRPIX with the PDD structure, in this simulation, $\Vbb$ and $\Vbpw$ were set to
$-300$~V and $-2.5$~V, respectively.

\begin{figure*}[htbp]
      \begin{center}
        \includegraphics[scale=0.45]{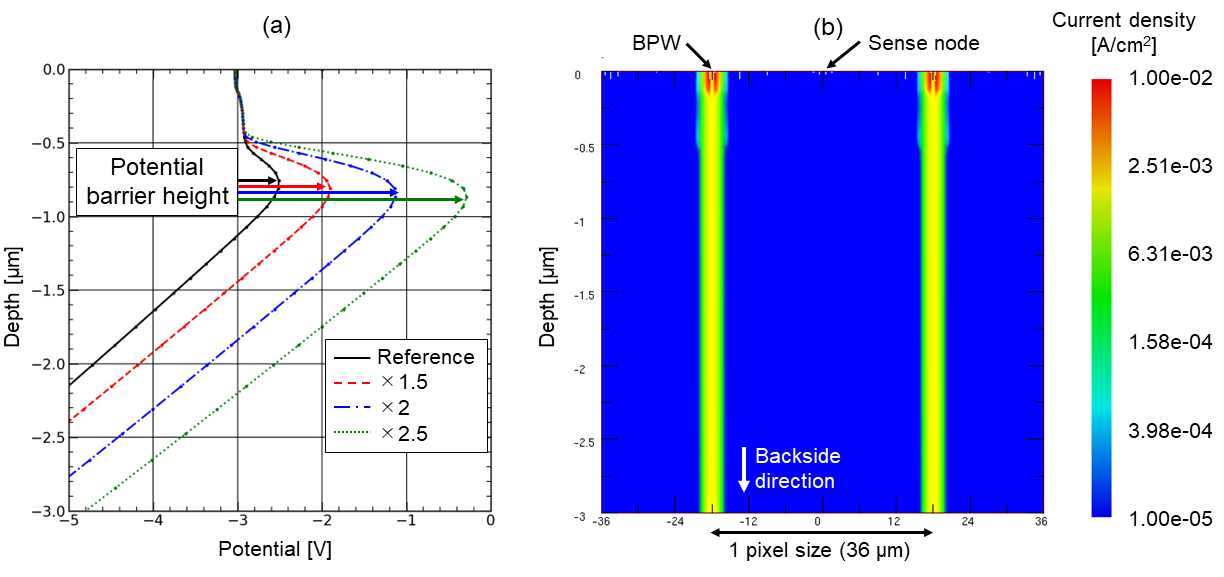}
        \caption{(a) Simulated potential profiles along line A--A' in Figure \ref{fig:schem_str} with several different dopant concentrations of BNW1. That with the reference value of the dopant concentration of BNW1 is shown in black solid line. Those with 1.5, 2.0, and 2.5 times the reference value are shown in red dashed, blue dash-dot and green dotted lines, respectively. The arrows correspond to the potential barrier height of BNW1. (b) Simulated current density map under the reference condition of BNW1.}
        \label{fig:sim_psijp}
      \end{center}
\end{figure*}

Figure \ref{fig:sim_psijp}(a) shows potential profiles along A--A' in Figure
\ref{fig:schem_str}, a vertical line from BPW to substrate crossing
BNW1. We performed simulations by increasing dopant concentrations of BNW1 from the
``reference'' value. Here, the reference value indicates the dopant
concentration of BNW1 of the previous prototype XRPIX with the PDD structure \cite{harada}.
The dopant concentration of BNW2 was constant at the reference value through the
simulations. The potential barrier height of BNW1 increases as the dopant
concentration increases. Figure~\ref{fig:sim_psijp}(b) shows a simulated current
density map under the reference condition of BNW1. A large leakage current from the
BPW to the substrate appeared because of punch-through. No leakage current appeared
in the other conditions. These results suggest that the potential barrier height
with the reference dopant concentration is insufficient to suppress punch-through.
On the other hand, as stated above, the dopant concentration of BNW1 cannot be
simply increased without consideration. The increase in dopant concentration would enlarge
the neutralized BNW1 region, leading to short circuit between neighboring sense nodes via BNW1.
The increase in dopant concentration would also narrow down the depletion region between the BNW1 and BPW, leading to an increase in sense-node capacitance.

\begin{figure}[htbp]
      \begin{center}
        \includegraphics[scale=0.5]{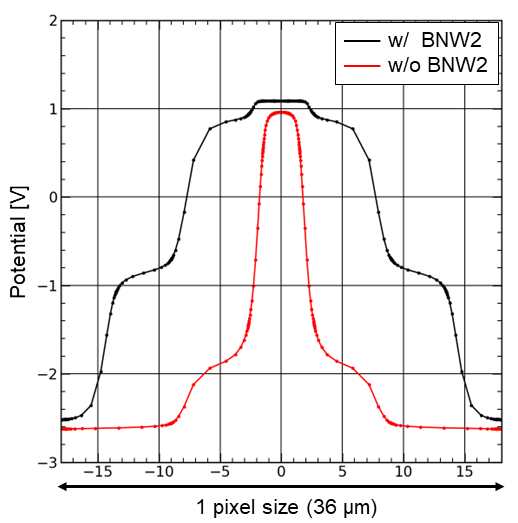}
        \caption{Simulated potential profile along line B--B' in Figure \ref{fig:schem_str} with BNW2 (black) and without BNW2 (red)}
        \label{fig:sim_Psi2}
      \end{center}
\end{figure}

We also performed simulations for the cases with or without BNW2.
Figure~\ref{fig:sim_Psi2} shows potential profiles along B--B' in
Figure~\ref{fig:schem_str}, a horizontal line crossing BNW2.
The potential shape near the pixel boundary is wider in the case with BNW2 compared to
that without BNW2. This indicates that BNW2 would help to increase the charge collection efficiency near the pixel boundary. On the other hand, the introduction of BNW2 results in an increase in sense-node capacitance.
It is also a subject of an optimization study between charge
collection efficiency and spectroscopic performance.

\subsection{Fabrication of candidate chips}\label{sec:fabrication}

\begin{figure}[htbp]
  \makeatletter
   \def\@captype{table}
   \makeatother
\begin{center}
  \caption{Specification of candidate chips}
  \label{tab:spec}
  \scalebox{0.8}[0.8]{
\begin{tabular}{ccc}
\hline\hline
Chip ID & BNW1: Dopant concentration & BNW2: Presence \\
\hline
XRPIX8.1 & Reference ($\times$1) & Yes \\
XRPIX8.2 & $\times$1.5 & Yes \\
XRPIX8.3 & $\times$2.0 & Yes \\
XRPIX8.4 & $\times$2.5 & Yes \\
XRPIX8.5 & $\times$2.0 & No \\
\hline
\end{tabular}
}
\end{center}
\end{figure}

In order to experimentally perform an optimization study on the n-well dopant concentration, we fabricated five candidate chips based on the TCAD simulation results.
All the chips were fabricated by using a 0.2~{\textmu}m
fully-depleted SOI CMOS pixel process prepared by Lapis Semiconductor Co.\
Ltd. Table~\ref{tab:spec} shows the specification of the chips. XRPIX8.1 has the
same dopant concentration of BNW1 as that of the reference value in TCAD simulation.
Punch-through between the BPW and the substrate was expected to
occur in XRPIX8.1 as predicted by the simulation result. From XRPIX8.2 to XRPIX8.4, the dopant
concentration of BNW1 is increased in order by a factor of 0.5 with respect to that of
XRPIX8.1. The dopant concentration of XRPIX8.2 was expected to be the lower limit that
can prevent punch-through set by the TCAD simulation. Although XRPIX8.4 has sufficient dopant
concentration to prevent punch-through, short circuit between the sense nodes via BNW1 might
occur. The dopant concentration of BNW2 is the same as the reference value of BNW1.
XRPIX8.5 has the same dopant concentration as that of XRPIX8.3 but does not have BNW2 for performance comparison.
Only XRPIX8.5 does not have BNW2 as shown in Table~\ref{tab:spec}.

\section{Experiments and results}\label{sec:exp}

\subsection{Basic characteristics}\label{sec:IV}

\begin{figure}[htbp]
      \begin{center}
        \includegraphics[scale=0.35]{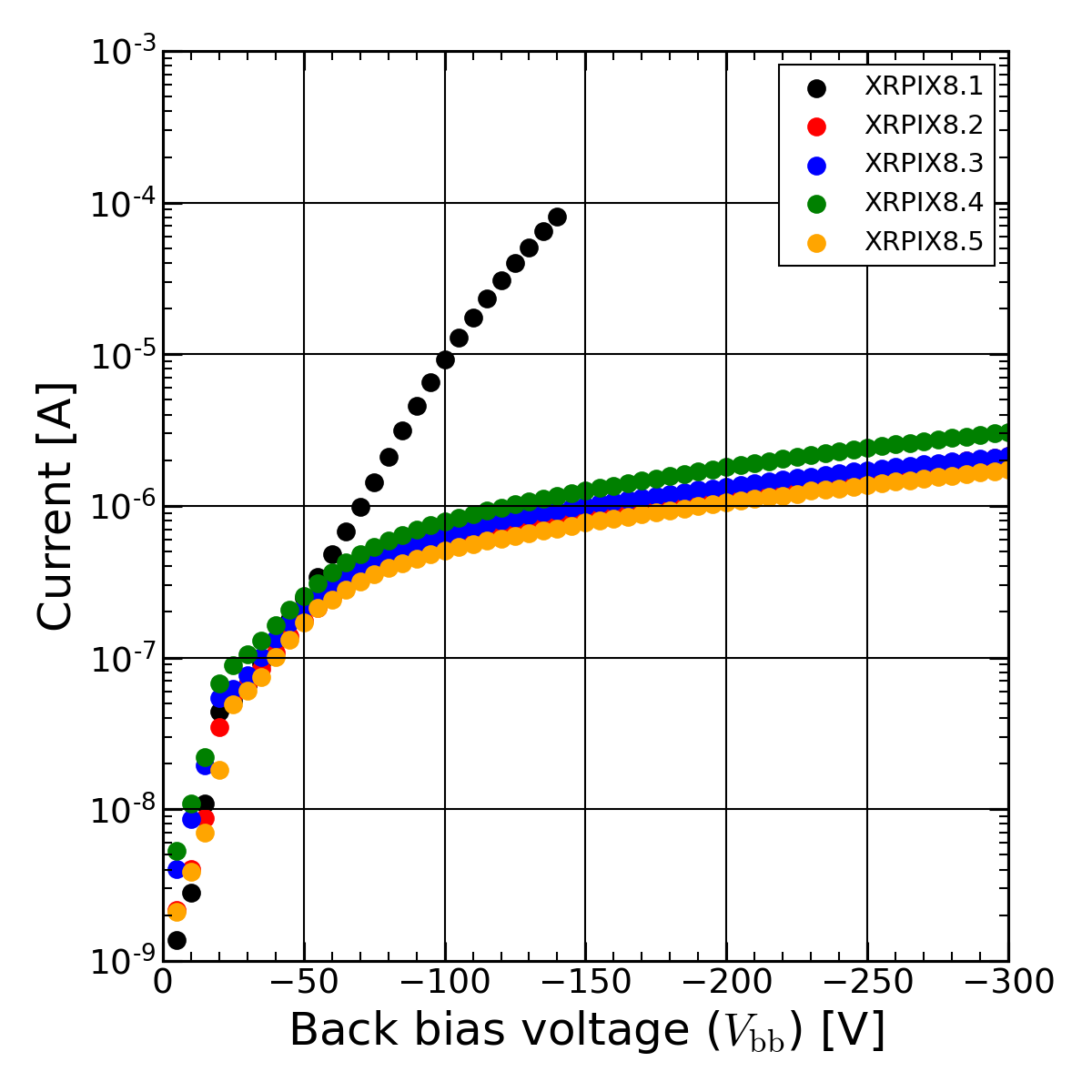}
        \caption{I-$\Vbb$ characteristics of the five candidate chips at room temperature.}
        \label{fig:exp_IV}
      \end{center}
\end{figure}

Figure \ref{fig:exp_IV} shows I-$\Vbb$ characteristics of the five candidate chips at room temperature. The I-$\Vbb$ characteristic of XRPIX8.1 deviates from others with $\Vbb$ greater than $-50$~V. This indicates that punch-through between the BPW and the substrate occurred due to a low potential barrier of XRPIX8.1 as suggested by our TCAD simulation. Other chips have breakdown voltages greater than $-300$~V, which is large enough for full depletion.
Results for leakage current suppression are summarized in Table~\ref{tab:performance}.

\begin{figure*}[htbp]
      \begin{center}
        \includegraphics[scale=0.5]{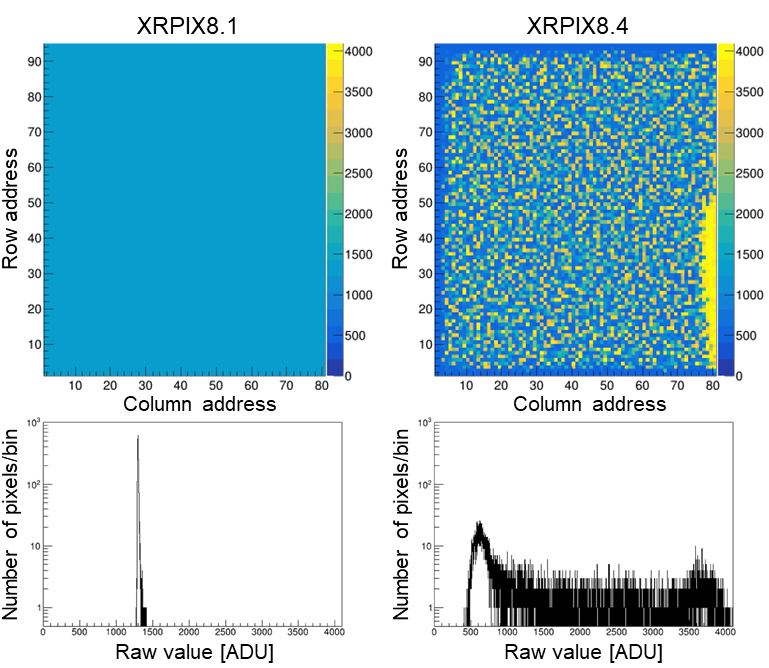}
        \caption{Raw-value maps and histograms in a single frame obtained with XRPIX8.1 (left) and XRPIX8.4 (right) with $\Vbb$ of $-5$~V at room temperature. These figures do not include the dummy pixels and the TEG-region pixels. The high raw-value region in the bottom right of the map of XRPIX8.4 is likely affected by the TEG-region pixels at column addresses greater than 80 in which circuit configurations are further different above and below row address of 48.}
        \label{fig:exp_cond}
      \end{center}
\end{figure*}

Figure \ref{fig:exp_cond} shows raw-value maps and histograms in a single frame obtained with XRPIX8.1
and XRPIX8.4 with $\Vbb$ of $-5$~V at room temperature.
Here, the raw value corresponds to the analog output of each pixel.
The chips were not irradiated with X-rays. The raw value of XRPIX8.4 significantly spatially
varies compared to that of XRPIX8.1. Other chips had spatial variations similar to XRPIX8.1.
The large spatial variation of raw value in XRPIX8.4 was improved by applying larger $\Vbb$,
which promotes the depletion of BNW1 under the BPW. This result suggests that
short circuit between the sense nodes via BNW1 occurs in the case of the high dopant
concentration of BNW1 at low $\Vbb$. It is not preferable that use of chip is limited with large $\Vbb$.
The dopant concentration of BNW1 is therefore desired to be less than 2.5 times
the reference value to prevent short circuit between the sense nodes via BNW1.

\subsection{Spectroscopic Performance}\label{sec:spec}

\begin{figure*}[htbp]
      \begin{center}
        \includegraphics[scale=0.45]{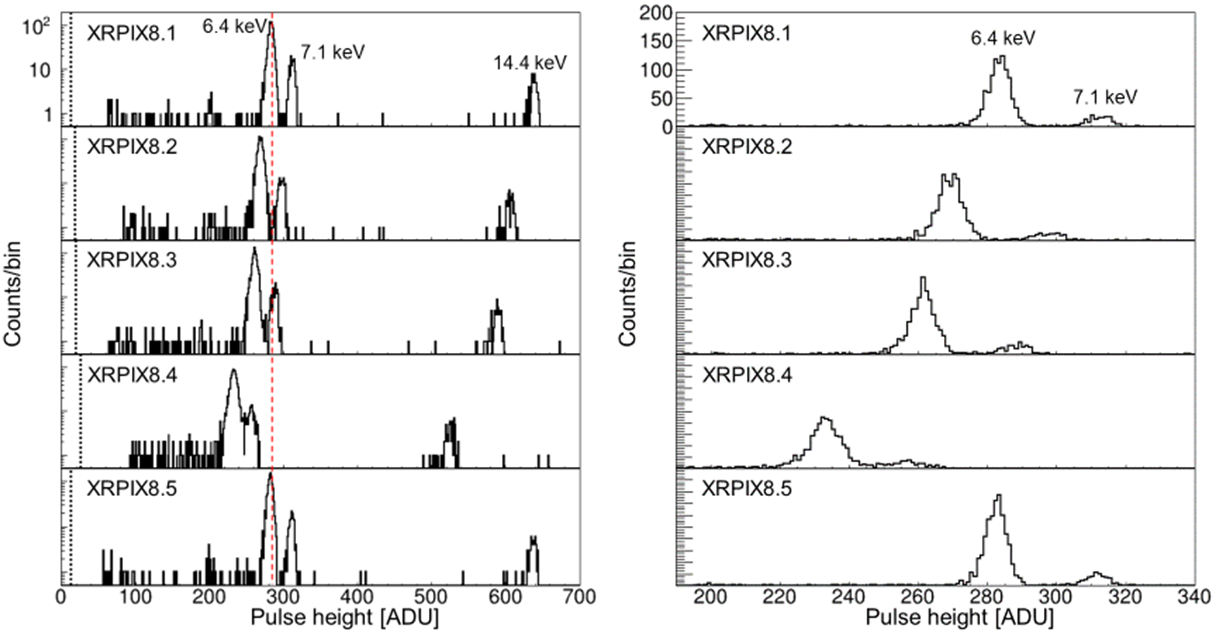}
        \caption{$^{57}$Co spectra obtained with the five candidate chips. The vertical axis in the left figure is a log scale to see the entire spectrum and the right figure is a linear scale to see a spectral line shape. Each spectrum was taken from one specific pixel and only single-pixel events were used. The black dotted line denotes the event threshold. The red dashed line denotes the line center of 6.4~keV in the spectrum of XRPIX8.1 for eye guide.}
        \label{fig:exp_spec}
      \end{center}
\end{figure*}

Figure~\ref{fig:exp_spec} shows $^{57}$Co spectra obtained with the five candidate
chips at $-60~^{\circ}$C.
In this experiment, the chips were irradiated from the front side with X-rays/$\gamma$-rays from $^{57}$Co.
Each spectrum was taken from one specific pixel that
showed the best performance among the pixels measured. In the data analysis, a
pedestal value in a given frame was calculated as a running average of 50 raw values
before and after the frame. A pedestal-subtracted-raw value was recorded as a
pulse height. Therefore, when no X-rays are detected, the pulse height histogram of the ``X-ray-non-detected pixels'' is a Gaussian-like distribution centered around zero, and the width of the distribution can be a measure of readout noise. We used the standard deviation of the Gaussian function fitted to the distribution to determine two thresholds. One threshold is the event threshold for X-ray detection and is set to 10 sigma. It corresponds to the lower energy threshold of the chips. The other is the split threshold for charge-sharing events and set to 3 sigma.
We used only single-pixel events, where signal charges
are collected within one pixel and no charge sharing with neighboring pixels occurs
with the split threshold. We applied $\Vbb$ of $-300$~V, which is large enough for
full depletion, to all the chips except for XRPIX8.1. Due to the large leakage
current caused by punch-through between the BPW and the substrate, $\Vbb$ applied
to XRPIX8.1 was limited to $-200$~V, which is still close to the full depletion
voltage. Three lines from $^{57}$Co (6.4, 7.1, and 14.4~keV) are seen in
each spectrum. The relationship between the pulse height, $\mathit{PH}$, and the
energy, $E$, of these lines are expressed as $\mathit{PH} = a E + b$. Each spectrum
is shifted by $b$ so that the line center in the unit of ADU simply indicates $a$,
namely overall gain. The amount of $b$ measured did not correlate with the chip
ID. From comparison among these spectra, it is clear that the line centers shift
toward lower pulse height from XRPIX8.1 to XRPIX8.4 and get back to the level
similar to XRPIX8.1 in XRPIX8.5. In addition, the line widths become larger from
XRPIX8.1 to XRPIX8.4 and again get back to the level similar to XRPIX8.1 in
XRPIX8.5.

\begin{figure*}[htbp]
  \makeatletter
   \def\@captype{table}
   \makeatother
\begin{center}
  \caption{Performance summary of the five candidate chips}
  \label{tab:performance}
\begin{tabular}{ccccc}
\hline\hline
Chip ID & Suppression of & Output gain & Energy resolution & Dark current \\
& leakage current & [{\textmu}V/e$^{-}$]  & at 6.4~keV (FWHM) [eV] & [ADU/ms] \\
\hline
XRPIX8.1 & No  & 39.42$\pm$0.06 & 166$\pm$5 & (2.46$\pm$0.09)~$\times~10^{-1}$ \\
XRPIX8.2 & Yes & 37.56$\pm$0.07 & 202$\pm$5 & (1.82$\pm$0.21)~$\times~10^{-1}$ \\
XRPIX8.3 & Yes & 36.43$\pm$0.07 & 208$\pm$6 & (2.76$\pm$0.40)~$\times~10^{-1}$ \\
XRPIX8.4 & Yes & 32.57$\pm$0.08 & 296$\pm$9 & (2.92$\pm$0.55)~$\times~10^{-1}$ \\
XRPIX8.5 & Yes & 39.38$\pm$0.05 & 166$\pm$4 & (4.35$\pm$0.11)~$\times~10^{-1}$ \\
\hline
\end{tabular}
\end{center}
\end{figure*}

Table~\ref{tab:performance} summarizes the output gain ($\Gout$) and energy resolution at 6.4~keV. $\Gout$ is calculated from $a$ in the equation shown above, using that 1~ADU corresponds to 448~{\textmu}V and that the mean ionization energy per electron-hole pair in silicon is 3.65~eV \cite{janesick}. The energy resolution represents FWHM of the 6.4~keV line. XRPIX8.1 and XRPIX8.5 show a satisfactory X-ray spectroscopic performance considering results obtained with previous XPPIX chips \cite{harada}. These numbers are actually good enough as a general silicon detector although not the best. On the other hand, as can be seen from the spectra in Figure~\ref{fig:exp_spec}, these performance gradually degrades from XRPIX8.1 to XRPIX8.4.

\begin{figure}[htbp]
      \begin{center}
        \includegraphics[scale=0.55]{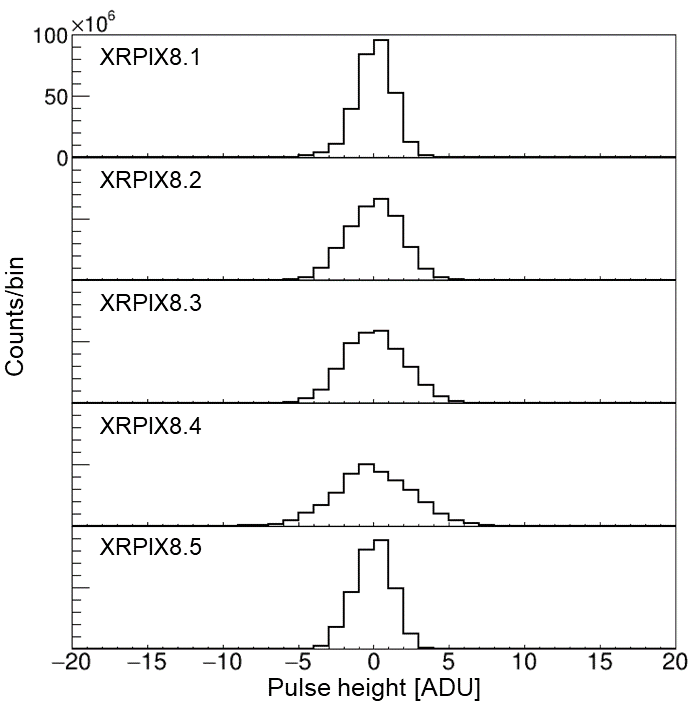}
        \caption{Pulse height distributions of X-ray-non-detected pixels of the five candidate chips.}
        \label{fig:exp_ron}
      \end{center}
\end{figure}

Increase in dark current and/or readout noise can result in the degradation of the
energy resolution. Dark current of each chip was measured following Kodama et al.
\cite{kodama}. The results measured at $-60~^{\circ}$C are also summarized in Table~\ref{tab:performance}. It
is indicated that the dark current does not correlate with chip ID unlike the energy
resolution and is not a major source of the degradation of the energy resolution.
Figure~\ref{fig:exp_ron} shows the pulse height distributions of X-ray-non-detected
pixels of the chips. The histogram width in Figure \ref{fig:exp_ron}
becomes larger from XRPIX8.1 to XRPIX8.4 and get back to the level similar to
XRPIX8.1 in XRPIX8.5. These results suggest that the increase in readout noise
mainly degrades the energy resolution.

All the five candidate chips except for XRPIX8.5 have BNW2, which is introduced for an improvement of charge collection efficiency. Insufficient charge collection efficiency typically results in a low-energy tail of spectral lines, which make a spectral line asymmetric \cite{kayama}. A comparison among the spectra in Figure~\ref{fig:exp_spec} indicates that the symmetry degrees of the lines do not change with or without BNW2. This result suggests that the introduction of BNW2 is not of critical importance in terms of charge collection efficiency.

Table~\ref{tab:performance} digests the experimental results presented here. We
found that the n-well design of XRPIX8.5 can make a
potential barrier high enough to suppress a leakage current and also provide
satisfactory X-ray spectroscopic performance. The detailed evaluation on the
spectroscopic performance of XRPIX8.5 is presented elsewhere.

\section{Discussion} \label{sec:discussion}

From the experimental results in section \ref{sec:spec}, we found that $\Gout$ and
the energy resolution gradually degrade from XRPIX8.1 to XRPIX8.4 and get back
to the level similar to XRPIX8.1 in XRPIX8.5. In this
section, we discuss this tendency in more detail.

\begin{figure}[htbp]
      \begin{center}
        \includegraphics[scale=0.35]{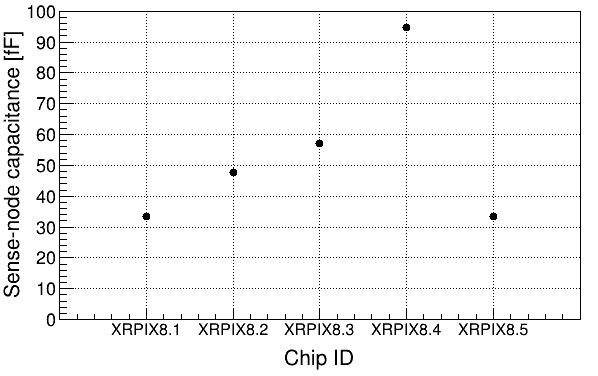}
        \caption{Sense-node capacitance ($\Cd$) calculated of the five candidate chips.}
        \label{fig:exp_gout2cd}
      \end{center}
\end{figure}

XRPIX has a charge-sensitive amplifier (CSA) and a source follower (SF) in a pixel circuit \cite{3b_csa}.
Therefore, $\Gout$ is a product of the CSA gain ($\Gcsa$), the SF gain ($\Gsf$), and the readout gain
($\Gro$) that is a resultant gain of the readout circuit of chip and instrumentation
amplifier on our evaluation board, which is written as
\begin{equation}
  \label{eq:gout}
\Gout = \Gcsa \times \Gsf \times \Gro.
\end{equation}
Here, $\Gsf$ was 0.95 from the circuit simulation and $\Gro$ was 0.70 from an experimental
measurement. $\Gcsa$ is defined as a ratio of output voltage to input charge in the
unit of elementary charge and is written as
\begin{equation}
  \label{eq:gcsa}
  \Gcsa = q\frac{\Acs}{\Cd+\Ci+(\Acs+1)\Cfb}.
\end{equation}
Here, $q$ is the elementary charge in coulomb, $\Cd$ is the sense-node capacitance,
$\Acs$, $\Ci$ and $\Cfb$ are open-loop gain, input capacitance, and feedback capacitance in the CSA, respectively.
$\Acs$ was designed to 108 \cite{DSOI}. $\Ci$ and $\Cfb$ were estimated to be 7.85~fF and 2.3~fF, respectively, by using the
parasitic capacitance extractor tool. Figure~\ref{fig:exp_gout2cd} shows
$\Cd$ of the five candidate chips, calculated using the $\Gout$ in Table~\ref{tab:performance}
and Equations~(\ref{eq:gout}) and (\ref{eq:gcsa}). $\Cd$ monotonically increases from XRPIX8.1
to XRPIX8.4 and decreases to the level similar to XRPIX8.1 in XRPIX8.5. This trend is naturally
interpreted as due to the variation in n-well dopant concentration around a sense node.
From XRPIX8.1 to XRPIX8.4, the dopant concentration of BNW1 increases monotonically
and that of BNW2 is constant, resulting in a monotonic increase of sense-node capacitance.
XRPIX8.5 does not have BNW2 and its dopant concentration of BNW1 is the same as that of XRPIX8.3,
resulting in the level of node capacitance similar to that of XRPIX8.1.

The relation between the energy resolution ($\Delta E_\mathrm{FWHM}$) and readout noise ($\sigma_\mathrm{R}$) is written as
\begin{equation}
  \label{eq:Eres}
  \Delta E_{\mathrm{FWHM}} = 2 \sqrt{2\ln 2} W \sqrt{\frac{FE}{W} + \sigma^{2}_{\mathrm{R}}}.
\end{equation}
Here, $E$, $F$, and $W$ are the incident X-ray energy, the Fano factor of 0.128
\cite{Kotov}, and the mean ionization energy per electron-hole pair in silicon of
3.65~eV, respectively. We first measured $\Delta E_\mathrm{FWHM}$ of the three lines
in Figure~\ref{fig:exp_spec} and derived $\sigma_\mathrm{R}$ by fitting the data of
``$\Delta E_\mathrm{FWHM}$ vs $E$'' with
equation~(\ref{eq:Eres}).

\begin{figure}[htbp]
      \begin{center}
        \includegraphics[scale=0.7]{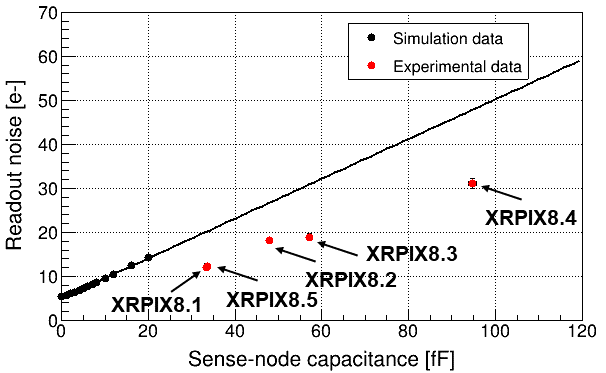}
        \caption{Readout noise ($\sigma_{R}$) as a function of the sense-node capacitance ($\Cd$). The red points and black points are experimental and simulation data, respectively. The black solid line is the fitting model (linear function) of simulation data.}
        \label{fig:sim_ron}
      \end{center}
\end{figure}

Figure~\ref{fig:sim_ron} plots $\sigma_{R}$ as a function
of $\Cd$. We also performed a circuit simulation to estimate $\sigma_{R}$, in which
we found that the flicker noise of the input transistor in the CSA almost dominates
the entire value. Figure~\ref{fig:sim_ron} also shows $\sigma_{R}$ as a function of
$\Cd$ obtained by the circuit simulation. The $\Cd$ dependencies of $\sigma_{R}$
obtained by our experiment and circuit simulation reasonably match, suggesting that
the variation of the energy resolution can be also explained by that of the sense-node
capacitance. On the other hand, there is a certain difference between the experiment
and simulation values. We consider that it might be due to the inaccuracy of $\Cfb$.
The feedback capacitance in the CSA largely depends on the Metal-Oxide-Metal
(MOM) capacitance. MOM capacitors employ parasitic capacitance between metal wires
so that the MOM capacitance is subject to small-scale manufacturing variations in
the layout and figure of the metal wires. Therefore, $\Cfb$ extracted from the
designed layout may not accurately reflect the real value. From
Equation~(\ref{eq:gcsa}), a small increase in $\Cfb$ leads to a large
decrease in $\Cd$, resulting in smaller differences between the experiment and
simulation values. We plan to implement TEG to measure $\Cfb$ experimentally in the
next chip.

\section{Conclusion} \label{sec:conculusion}
An optimization study on the n-well dopant concentration in the
XRPIX with a PDD structure was performed.
Five candidate chips were fabricated based on the TCAD simulation, focusing on the potential
barrier height between the pinned p-well and p-type substrate and also on the
sense-node capacitance. The best n-well design was successfully found out in which
a large leakage current was surely suppressed and satisfactory X-ray spectroscopic performance
was obtained. The X-ray spectroscopic performance actually depends on the dopant concentration and
configuration of the n-well, which can be basically understood by the variation in sense node capacitance.

\section{Acknowledgments}
We acknowledge the valuable advice and the manufactures of XRPIXs by the personnel of LAPIS Semiconductor Co., Ltd.
This study was supported by the Japan Society for the Promotion of Science (JSPS) KAKENHI, Grant-in-Aid for Scientific Research on Innovative Areas 25109004 (T.G.T., T.T., K.M., A.T., and T.K.),
Grant-in-Aid for Scientific Research (A) 15H02090 (T.G.T. and A.T.) and 21H04493 (T.G.T., T.T. and A.T.), Grant-in-Aid for Scientific Research (B) 21H01095 (K.M.), 22H01269 (T.K. and K.H.) and by JST SPRING, Grant Number JPMJSP2105.
This study was also supported by the VLSI Design and Education Center (VDEC), the University of
Tokyo in collaboration with Cadence Design Systems, Inc., and Mentor Graphics, Inc.

\bibliography{mybibfile}

\end{document}